\documentclass[twocolumn,english,aps]{revtex4-2}
\usepackage[T1]{fontenc}
\usepackage[latin9]{inputenc}
\usepackage{color}
\usepackage{amsmath}
\usepackage{graphicx}
\usepackage{comment}
\usepackage{sidecap}
\usepackage[colorlinks=true, allcolors=red]{hyperref}

\makeatletter

\usepackage{babel}

\makeatother

\usepackage{babel}
\begin{document}
\title{Merging dynamics of plasma blobs in the Scrape-off Layer of a tokamak}
\author{Souvik Mondal$^{1,3}$, N Bisai$^{1,3}$, Abhijit Sen$^{1,3}$ and Indranil Bandyopadhyay$^{1,2,3}$}
%\email{E-mail: souvik.mondal@ipr.res.in}
\affiliation{$^1$Institute for Plasma Research, Bhat, Gandhinagar 382428, Gujarat, India \\
$^2$ITER-India, Institute for Plasma Research, Bhat, Gandhinagar 382428, Gujarat, India \\
$^3$Homi Bhabha National Institute, Training School Complex, Anushaktinagar, Mumbai 40094, India}

%------------------------------------------------------------------------------------------------------------------------------------------------------------------%
\begin{abstract}
%------------------------------------------------------------------------------------------------------------------------------------------------------------------%
The emergence and merging of high-density coherent structures - plasma blobs - is a recurrent phenomenon in the Scrape-off layer (SOL) of a tokamak plasma that has a significant impact on the rate of convective transport in that region. We report on a model study of the merging of two electromagnetically interacting blobs in a high beta plasma. Our detailed numerical simulations show that the merging process is akin to the coalescence instability between two magnetic islands but with important differences due to the density perturbation. The blobs are found to rotate about each other during merging and the merging occurs with an acceleration in the poloidal direction that is directly proportional to the square of the current density of the blobs and inversely proportional to its density. The separation distance between two high current density blobs is also seen to oscillate indicating a sloshing behavior. 
\end{abstract}

\maketitle

%------------------------------------------------------------------------------------------------------------------------------------------------------------------%
\section{INTRODUCTION}
%------------------------------------------------------------------------------------------------------------------------------------------------------------------%

Several past studies have demonstrated that intermittent and fast convective transport that takes place in the edge and scrape-off layer (SOL) regions of a tokamak plasma can be attributed to the radial motion of coherent density structures called ``blobs'' \cite{sarazin_intermittent_1998, Umansky_1998, KRASHENINNIKOV2001368,Zweben_2002,Boedo_2002,J_L_Terry_2003,Bisai_2004, bisai_3f2f_2004,Bisai_2005,Shankar_2021,bisai_blob_2022,bisai_experimental_2022, Zweben_2022,bisai_rmpp_2023}. Blobs are nonlinear filamentary structures that are extended along the direction of the magnetic field and have a smaller spatial extent in the direction perpendicular to the magnetic field. They have a higher density and temperature than the surrounding plasma and also contain a parallel current density. The intrinsic current density can be of a ``dipole'' nature (currents flowing in the parallel and anti-parallel directions) or have a ``mono-pole'' nature (a unidirectional current) \cite{myra_current_carrying_filament_2007}. Blobs are generated non-linearly from the interchange/drift plasma turbulence in the edge/SOL region. They can further interact with each other leading to either their breakup or a coalescence into a larger blob. Since convective plasma transport is directly proportional to the size (mass) of the coherent structure, the blob-merging process can significantly impact the rate of plasma transport. As the plasma turbulence in the edge/SOL region can be electromagnetic for high temperature and high beta ($\beta$) plasmas, the increase of the magnetic fluctuations by the merging processes can also modify the state of the electromagnetic turbulence. The above significant consequences of blob merging, namely their influence on plasma transport and turbulence, underscore the importance of investigating the detailed dynamics of blob merging and are the primary motivation of the present study.\\

As charged objects, blobs can influence each other's dynamics through their electrostatic interaction. The nature of the electrostatic interaction has been shown to be very small unless their separation is less than five times their width \cite{Militello_2017_Blob_interactions_in_2D_seeded}. The electrostatic interaction between a small blob and a larger blob moving from the same poloidal location in the radial direction has been studied by D'Ippolito \textit{et. al} \cite{D'Ippolito_2003} to investigate the stability of the blob against fragmentation. Blobs can also interact with each other electromagnetically through the Lorentz force on account of their intrinsic parallel currents \cite{myra_current_carrying_filament_2007}. This force can be attractive if the direction of the parallel current is the same in the two blobs and this can lead to their eventual merger. In this work, we study such a case of two blobs with intrinsic initial ``monopolar'' currents in the same direction as they dynamically merge into each other.  For simplicity, we neglect the radial motion of the blobs that can arise from magnetic curvature-induced charging of the blob in the poloidal ($y$) direction that generates a poloidal electric field $E_y$ and causes a radial motion by the $\vec{E}_y\times \vec{B}$ drift. In our model, we are interested in the blob merging taking place in the poloidal direction. The underlying physical mechanism leading to the merger is very akin to that governing the coalescence instability that has been widely studied in the context of merging of magnetic islands \cite{Coalescence_instability_finn_kaw, Pritchett_1979, Biskamp_PRL_1980, biskamp_PRA_1982, Bhattacharjee_1983, Rickard_MR_1993, Jagannath_pop_2021}.\\

For our present simulations, we have used SOL plasma parameters that are close to high-$\beta$ ITER discharge scenarios. Two blobs with 2D bi-Gaussian profiles in the ($x,y$) plane that are poloidally separated by a distance are seeded for the merging study. Each blob is assumed to carry a high parallel current density $J_{\|0}\sim (0.1-1)$ MA/m$^{2}$. Two sets of studies are done - one without the perturbed density profile of the blob ($n_b = 0 $, where $n_b$ denotes an initial amplitude of density perturbation) and one in the presence of such a profile ($n_b \neq 0$). The merging of the two blobs has been studied as a function of the quantities like blob size ($\delta$), $n_b$, and $J_{\|0}$. \\

This paper is organized as follows. Section \ref{sec:Model_Equations} presents the model equations for the blob dynamics followed by some qualitative insights of the blobs in Section \ref{sec:qualitative insights}. The details of the numerical simulation, which includes the input parameters, initial conditions, and boundary conditions, are provided in Section \ref{sec:Numerical_simulation}. The simulation results are detailed in section \ref{sec:simulation results}. Section \ref{sec:discussion} provides a brief summary and some concluding remarks.

%------------------------------------------------------------------------------------------------------------------------------------------------------------------%
\section{MODEL EQUATIONS\label{sec:Model_Equations}}
%-----------------------------------------------------------------------------------------------------------------------------------------------------------------%
Our model equations are based on the Braginskii equations \cite{braginskii1965transport} in a three-dimensional (3D) Cartesian geometry with the magnetic field directed along the $z$-direction and $x$,$y$ representing the radial and poloidal directions, respectively. To physically represent the SOL - the physical space in the $x,y$ plane where the magnetic field terminates at the material plates representing the limiter or the divertor. The electron and ion temperatures are assumed to be uniform with $T_e\gg T_i$ (cold ion approximation). The plasma beta ($\beta=8\pi nT_e/B_0^2$) is assumed to be sufficiently high so that the inductive part of the electric field $E_{\|}^{ind}=-1/c(\partial A_{\|}/\partial t)$ cannot be neglected and electromagnetic effects are retained.
The basic model equations consist of the conservation equations for the plasma density and plasma current (or quasi-neutrality), generalized Ohm's law, and Maxwell's equations. We further assume the magnetic field to be uniform so that the magnetic curvature-induced effective gravity force is absent. Under the above approximations, the model set of equations is given as \cite{lee_electromagnetic_2015, stepanenko_impact_2020} :

\begin{equation}
	\frac{e\rho^2_{s}}{T_{e}}n\frac{d}{dt}(\nabla_{\perp}^{2}\phi)=\frac{1}{e}\nabla_{\|}J_{\|},\label{eq:unnormalized vorticity equation}
\end{equation}
\begin{equation}
\frac{dn}{dt}=\frac{1}{e}\nabla_{\|}J_{\|},\label{eq:unnormalized continuity equation}
\end{equation}
\begin{equation}
-\frac{e}{m_e c}\frac{dA_{\|}}{dt}=\frac{e}{m_{e}}\frac{\partial\phi}{\partial z}-\frac{T_{e}}{m_{e}}\nabla_{\|}\ln n+\frac{e}{\sigma_{\|}m_{e}}J_{\|}.\label{eq:vector potential equation}
\end{equation}

Equations - (\ref{eq:unnormalized vorticity equation}), (\ref{eq:unnormalized continuity equation}), and (\ref{eq:vector potential equation}) represent current conservation, density continuity, and generalized Ohm's law. Here $n$ is the plasma density, $\phi$ is the electrostatics potential, $\omega=\nabla_{\perp}^2\phi$ is the vorticity, $A_{\|}$ is the parallel magnetic vector potential, $c$  is the speed of light, and $m_e$, $T_e$, and $e$ are  mass, temperature, and charge of the electron, respectively. $\rho_s=c_s/\Omega_s$ is the ion gyro-radius where $c_s=\sqrt{T_e/m_i}$ is the sound speed, $\Omega_s$ is the gyro-frequency, and $m_i$ is the ion mass. The parallel electrical conductivity is $\sigma_{\|}=1.96n e^{2}/m_{e}\nu_{ie}$, where $\nu_{ei}=2.9\times10^{-6}n\ln\Lambda/T_{e}^{3/2}$ is the electron-ion collision frequency with $\ln\Lambda \simeq 10$ representing the Coulomb logarithm. The total time and space derivative operators are defined as ${d}/{dt}={\partial}/{\partial t}+({c}/{B_0})\hat{b}_0\times \nabla\phi \cdot \nabla$ and $\nabla_{\parallel}= {\partial}/{\partial z} + ({\nabla A_\parallel}/{B})\times \hat{b}_0 \cdot \nabla$, where $\hat{b}_0$ is the unit vector along the unperturbed magnetic field. From Ampere's Law, the parallel current density can be expressed in terms of the parallel vector potential as::
\begin{equation}
J_\parallel=-\frac{c}{4\pi}\nabla_{\perp}^2A_\parallel.
\end{equation} 
Since the perpendicular advection time $\tau_{\perp}\sim \delta/v_{E\times B}$ of a blob is much larger than the electron-ion collision time ($\tau_{ei}$), the electron inertia term can be neglected  \cite{stepanenko_impact_2020}.
We further simplify Eqs.(\ref{eq:unnormalized vorticity equation})-(\ref{eq:vector potential equation}) to a 2D geometry by  averaging along the magnetic field direction $(\nabla_{\|}\sim 1/L_{\|})$, where $L_{\|}$ is the parallel connection length. Using the inductive part of the parallel derivative term $({\nabla A_\parallel}/{B_0})\times \hat{b}_0 \cdot \nabla$ in Eqs. (\ref{eq:unnormalized vorticity equation}) and (\ref{eq:unnormalized continuity equation}) the parallel derivative-related term can be written as
\begin{equation}
    \nabla_{\|}J_{\|}=\frac{\partial J_{\|}}{\partial z}+({\nabla A_\parallel}/{B_0})\times \hat{b}_0 \cdot \nabla J_{\|}. \label{eq:pararllel_approx}
\end{equation}
The second term of Eq.(\ref{eq:pararllel_approx}) can also be written in a Poisson bracket form, i.e. $({\nabla A_\parallel}/{B_0})\times \hat{b_0} \cdot \nabla)f=(1/B_0)[f,A_{\|}]_{x,y}$. In the 2D model approximation, we keep this Poisson bracket term and neglect the parallel term as $\partial/\partial z=0$. Consequently, the parallel derivative term will be $\nabla_{\|}J_{\|}=(1/B_0)[J_{\|},A_{\|}]_{x,y}$ and $\nabla_{\|}\ln n=(1/B_0)[\ln n,A_{\|}]_{x,y}$.\\

To numerically simulate 
Eqs.(\ref{eq:unnormalized vorticity equation})-(\ref{eq:vector potential equation}), we use the following normalizations:
${n}/{n_{0}}\rightarrow n$, $t\rightarrow t\Omega_{s}$, ${(x,y)}/{\rho_{s}}\rightarrow (x,y)$, $v_{\parallel}/{c_{s}}\rightarrow v_{\parallel}$, ${J_{\parallel}}/{n_{0}ec_{s}}\rightarrow J_{\parallel}$, ${e\phi}/{T_{e0}}\rightarrow\phi$, ${A}/{B_{0}\rho_{s}}\rightarrow A$, ${T_{e}}/{T_{e0}}\rightarrow T_e$ and $L_{\parallel}{\nabla_{\parallel}}\rightarrow\nabla_{\parallel}$ where $n_0$, $T_{e0}$, and $B_0$ represent the values of plasma density, electron temperature, and toroidal magnetic field  in Last Closed Flux Surface (LCFS). $v_\|$ represents the parallel velocity of electrons.\\

Using the above approximations and normalizations the final form of the equations can be expressed as :
\begin{equation}
\frac{\partial n}{\partial t}=-[\phi,n]-\left[A_{\|},J_{\|}\right],\label{eq:norm_density_continuty_eq}
\end{equation}
\begin{equation}
n\frac{\partial}{\partial t}(\nabla_{\perp}^{2}\phi)=-[\phi,\nabla_{\perp}^{2}\phi]-\left[A_{\|},J_{\|}\right],\label{eq:norm_vorticity_eq}
\end{equation}
\begin{equation}
\frac{\partial A_{\|}}{\partial t}=-\left[\phi,A_{\|}\right]-\left[A_{\|},\ln n\right]+\eta \nabla_{\perp}^{2}A_{\|},\label{eq:norm_vector_pot_eq}
\end{equation}
and 
\begin{equation}
J_\parallel=-a\nabla_{\perp}^2A_\parallel,\label{eq:norm_maxwell_eq}
\end{equation}
where $\eta=1/(\Omega_s \tau_s)$ is the normalized magnetic diffusion constant, $\tau_{s}=4\pi\sigma_{\|}\rho_{s}^{2}/c^{2}$ is the magnetic screen time, and $a=  1.96{m_{i}}/{m_{e}\nu_{ie}\tau_{s}}$.

%------------------------------------------------------------------------------------------------------------------------------------------------------------------%
\section{Some analytical insights into blob dynamics}
\label{sec:qualitative insights}
%------------------------------------------------------------------------------------------------------------------------------------------------------------------%

\noindent Before carrying out a numerical simulation of the model equations described in the previous section, it is worthwhile to obtain some analytic insights that will be helpful in the interpretation and understanding of the numerical results. We begin by rewriting the equations in the $E\times B $ frame as 

\begin{equation}
	\frac{dn}{dt}=-\left[A_{\|},J_{\|}\right],\label{eq:1}
\end{equation}
\begin{equation}
	n\frac{d}{dt}(\nabla_{\perp}^{2}\phi)=-\left[A_{\|},J_{\|}\right],\label{eq:2}
\end{equation}
\begin{equation}
	\frac{d}{dt} A_{\|}=-\left[A_{\|},\ln n\right]+\eta \nabla_{\perp}^{2}A_{\|}.
\end{equation}
From Eq.(\ref{eq:1}) and (\ref{eq:2}) we can write,
\begin{equation}
	n\frac{d}{dt}(\nabla_{\perp}^{2}\phi)=\frac{dn}{dt},
\end{equation}
integrating both sides we get,
\begin{equation}
	 \nabla_{\perp}^{2}\phi=\ln\left(\frac{n}{n_{t=0}}\right) \label{eq:ln_condition}
\end{equation}

which shows that plasma vorticity is a function of density. $n_{t=0}$ indicates initial plasma density.  The plasma blob has a high density at the center and falls radially outward (as we assume a Gaussian profile), therefore $n$ is a function of the blob's radius $r$. If Eq.(\ref{eq:ln_condition}) is solved to determine $\phi$, it will be a function of $r$.  This will give a radial electric field $\vec{E_r}$ that will cause a poloidal rotation/spin by $E_r\times B$ drift in the azimuth direction. Therefore, in the presence of a blob density profile, the blob will rotate/spin as per $E_r$ obtained from the solution of $\phi$ from Eq.(\ref{eq:ln_condition}). It is to be noted that the origin of the rotation/spin is different from the conventional blob rotation in the SOL when it is connected to the sheath \cite{myra_2004} as that $\phi\propto T_e(r)$. Here, the rotation/spin occurs as the polarization current is balanced by $\nabla_\| J_\|$ as in Eq.(\ref{eq:2}). The spin of the blob will be discussed further in section-\ref{sec:simulation results}.\\

We next estimate the acceleration towards each other when two blobs experience an attraction due to a Lorentz force. Eq. (\ref{eq:unnormalized vorticity equation}) can be written as 
\begin{equation}
	n\frac{d}{dt}(\nabla_{\perp}^{2}\phi)=\hat{z}\cdot \vec{\nabla}_{\perp} J_\|\times \vec{\nabla}_\perp A_\|,\label{eq:A1}
\end{equation}
From Ampere's law, for two blobs of perpendicular radius $\delta_b$, separated by a distance $d$,  we can then write in the normalized form $\nabla_\perp A_{\|}=2\pi n_0ec_s J_{\|}\delta^2/dcB_0$ \cite{myra_current_carrying_filament_2007}, and also using $\nabla_\perp \sim 1/\delta$ we can write 
\begin{equation}
	n\frac{d\phi}{dt}=\frac{2\pi n_0ec_s \delta^3}{B_0dc\rho_s}J^2_{\|},\label{eq:A3}
\end{equation}
Using the steady blob velocity $v\sim cT_e\phi/\delta_eB_0c_s$, due to the magnetostatic forces between the filaments, the acceleration of the filaments can be written as 
\begin{equation}
    \frac{dv}{dt}=\frac{2\pi n_0m_ic^2_s \delta^2}{d\rho_sB^2_0}\frac{J^2_{\|}}{n},\label{eq:A4}
\end{equation}
Finally, after some calculation, the acceleration of the filaments can be written as
\begin{equation}
    \frac{dv}{dt}=f\frac{J^2_{\|}}{n},\label{eq:A5}
\end{equation}
where $f=\delta^2(m_e/m_i)(\tau_s \nu_{ei})/4d\rho_s$ is the normalization constant with $\nu_{ei}$ and $\tau_s$ as defined in the previous section.\\

Thus the acceleration between two blobs is proportional to the square of the initial $J_\|$ and inversely proportional to $n$. In Section-\ref{sec:simulation results} we will compare this estimate with the values of the acceleration obtained from the numerical simulation. 

%------------------------------------------------------------------------------------------------------------------------------------------------------------------%
\section{NUMERICAL SIMULATION\label{sec:Numerical_simulation}}
%------------------------------------------------------------------------------------------------------------------------------------------------------------------%

The numerical simulation of Eqs. (\ref{eq:norm_density_continuty_eq})-(\ref{eq:norm_maxwell_eq}) is done based on the input parameters related to high $\beta$ plasma of ITER tokamak that are given in Table (\ref{table:1}).

\begin{table}[h!]
\centering
\setlength{\tabcolsep}{12pt}
\begin{tabular}{l c c} 
 \hline
 Parameter & Value & Unit \\ [0.5ex] 
 \hline\hline
 $n_0$ & $1\times10^{14}$ & cm$^{-3}$ \\ 
 $T_e$ & 200 & $eV$ \\
 $B_0$ & 5.3 & T \\
 $R$ & 6 & m \\
 $L$ & 100 & m \\
 $c_s$ & 9.98$\times10^{6}$ & cm/sec \\ 
 $\Omega_s$ & 2.64$\times10^{8}$ & /sec \\
 $\rho_s$ & 3.78$\times10^{-2}$ & cm \\
 $\nu_{ei}$ & $1.02\times10^{6}$ & /sec \\
 $\sigma_{\|}$ & $4.82\times10^{16}$ & /ohm-cm \\
 $\delta$ & 0.7 & cm \\ [1ex] 
 \hline
\end{tabular}
\caption{Typical input plasma parameters. For the simulation, plasma parameters near LCFS are used.}
\label{table:1}
\end{table}

\noindent Initially at $t=0$, in our simulation, we seed two blobs in Gaussian form in the perpendicular plane centered at different poloidal locations $y_1$ and $y_2$ as:
\begin{equation}
n(\textbf{r},0) = 1+n_b \exp\left[-\frac{(x-x_0)^2[(y-y_1)^2+(y-y_2)^2]}{\delta^2}\right],
\end{equation}
where $x_0=L_x/2$ is the initial position of the center of mass of the blob in the radial direction, $y_1=0.3L_y$ and $y_2=0.7L_y$. $L_x=L_y=256\rho_s$ are lengths of the simulation domain in the radial and poloidal directions, respectively. $n_b$ is the initial relative blob amplitude. $\delta$ is the cross-sectional size of the blob.  The corresponding equilibrium parallel current densities $J_{\|}$ are taken to be in the same form of density perturbation in the perpendicular plane\cite{myra_current_carrying_filament_2007} as $$J_{\|} = J_{0}\exp\left[-\frac{(x-x_0)^2[(y-y_1)^2+(y-y_2)^2]}{\delta^2}\right],$$
where $J_{0}$ is the amplitude of equilibrium parallel current density. \\

In our simulation, we consider two cases: (i) the presence of the blob density profile (BDP) ($n_b\neq 0$) and (ii) the absence of the blob density profile ($n_b=0$). The equilibrium profile of parallel magnetic vector potential $A_{\|}$ is estimated from Eq.(\ref{eq:norm_maxwell_eq}) numerically using Laplace inversion at the initial time. In this simulation, the initial conditions of other ($\nabla_\perp^2 \phi$ and $\phi$) variables are zero. In solving our model equations, we set periodic boundary conditions in the $y$-direction and Dirichlet boundary conditions in the $x$-direction for $n$, $\phi$, $\nabla_\perp^2\phi$, $J_{\|}$, and $A_{\|}$.\\

The equations are solved numerically using the BOUT++ framework \cite{DUDSON_bout++}, where for space derivatives, we use a fourth-order central difference scheme in $x$ and FFT-based calculations in $y$ directions, along with a third-order WENO scheme for the upwind derivatives. For time integration, we used the CVODE solver. The number of grids used in our model has dimensions $N_x\times N_y=516\times 512$ in the $x$ and $y$ directions respectively that give minimum grid sizes of $dx=dy=0.5\rho_s$.

\section{Simulation results} \label{sec:simulation results}
The numerical simulation results for the merging process of two poloidally separated plasma blobs of the same size will be presented here. Unless otherwise mentioned all the quantities have both the equilibrium and fluctuation components and all are mentioned in the normalized units.
%------------------------------------------------------------------------------------------------------------------------------------------------------------------% 
\subsection{Merging in the presence and absence of blob density profile}
%------------------------------------------------------------------------------------------------------------------------------------------------------------------% 
\begin{figure*}
    \centering
    \includegraphics[width=0.65 \linewidth]{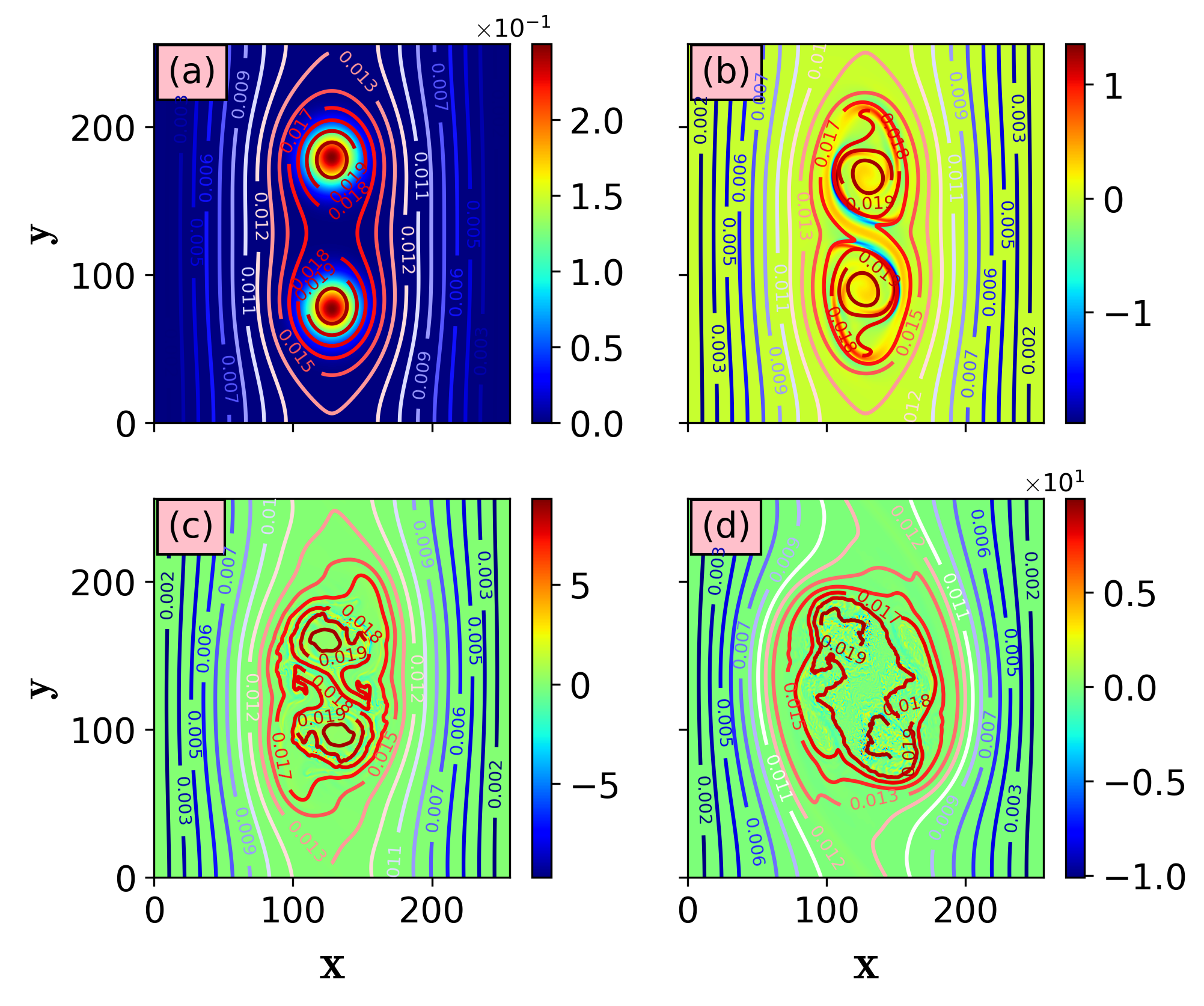}
    \caption{The time evolution of $A_\|$(contour) and $J_\|$ (color) in the case of the presence of blob density profile [(a)-(d)] for the amplitude of initial current density $J_0=$ 0.25, and two $J_0$s are separated at a distance of $102\rho_s$. Here (a) $t=10/\Omega_s$, (b) $t=4000/\Omega_s$, (c) $t=8000/\Omega_s$, and (d) $t=12000/\Omega_s$.}
    \label{fig:3f_ap_jp}
\end{figure*}

\begin{figure}
    \includegraphics[width=0.99 \linewidth]{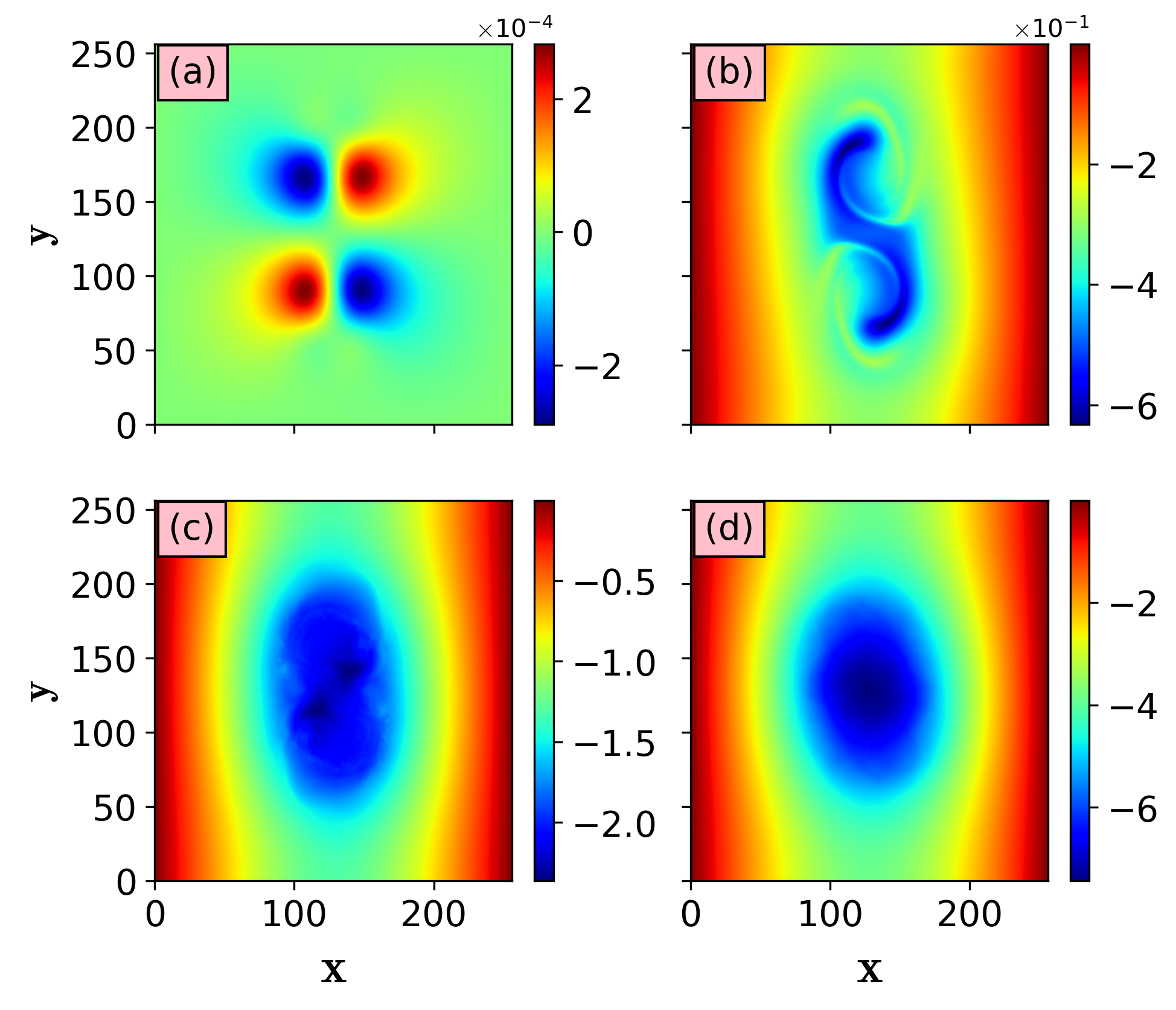}
    \caption{The time evolution of electrostatic potential in the case of blob density profile. Here, the amplitude of initial current density $J_0=$ 0.25, and two $J_0$s are separated at a distance of $102\rho_s$. Here (a) $t=10/\Omega_s$, (b) $t=4000/\Omega_s$, (c) $t=8000/\Omega_s$, and (d) $t=12000/\Omega_s$.}
    \label{fig:phi_3f}     
\end{figure}
Here, we will present numerical results obtained from solving Eqs.(\ref{eq:norm_density_continuty_eq}),(\ref{eq:norm_vorticity_eq}), and (\ref{eq:norm_vector_pot_eq}) with the presence of blob density profile where the current density was present initially and is derived from the initial $n$ so that $J_0=en(V_{\|i}-V_{\|e})$ where $V_{\|i}$ and $V_{\|e}$ are parallel velocity of ions and electrons; $V_{\|i}\sim c_s$. Figure \ref{fig:3f_ap_jp} illustrates the snapshot of the superposition of $A_\|$ (contour) and $J_\|$ (color). Initially, two identical Gaussian-type parallel currents having amplitude $J_0=$ 0.25 ($\sim 0.4$ MA/m$^2$) are separated by a distance of $102\rho_s$ in the poloidal direction. The attraction force between the two parallel currents gives an unstable state and makes them come closer to finding a stable configuration. We observe that the magnetic flux surfaces tend to coalesce and finally merge due to the onset of $E_x$ and $-E_x$ that provides $E_x\times B$ drifts to come closer. \\

It is also observed that in the course of merging the blobs rotate around a local axis where $J_0$ is maximum. This rotation of the filaments can be understood from Eq.(\ref{eq:ln_condition}) as discussed in Sec-\ref{sec:qualitative insights}. This condition indicates that $\omega$ is a logarithmic function of the plasma density. So, in the presence of a blob density profile, $\omega$ is also $r$-dependent. Due to this $\phi$ also depends on the local profile of the blob density structure. The snapshots of $\phi$ are shown in Fig.\ref{fig:phi_3f}(a)-(d). It has been observed that in the early time of evolution, the structure of $\phi$ is a quadrupole-like structure (see Fig.\ref{fig:phi_3f}(a)), as it evolves further the structure becomes monopolar as seen in Figs.\ref{fig:3f_ap_jp}(b)-(d) and its magnitude increases with time. This monopolar structure in the potential generates the rotation of the filaments \cite{Bisai_2004}. \\

\begin{figure*}
    \includegraphics[width=0.65 \linewidth]{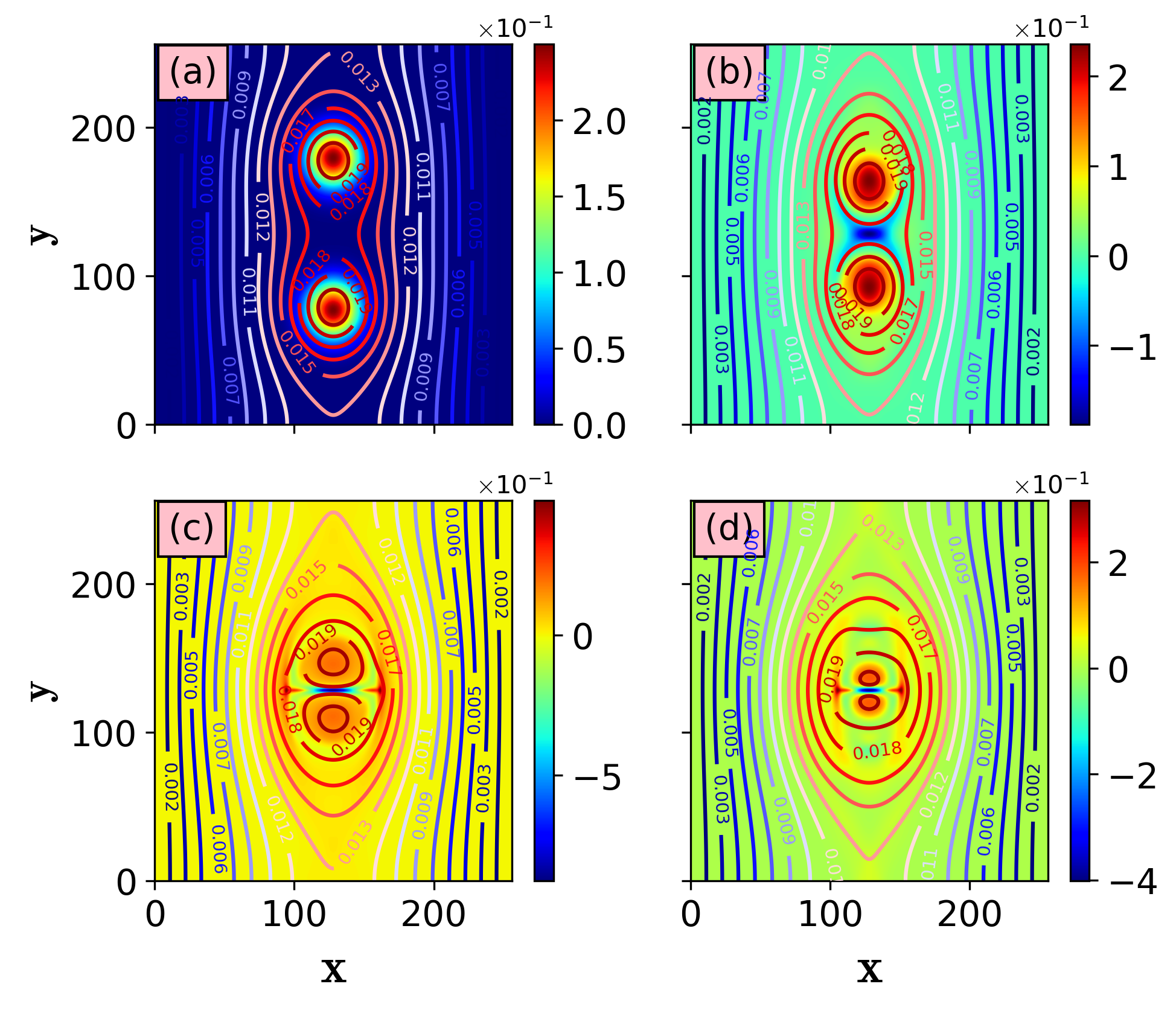}
\caption{The time evolution of $A_\|$(contour) and $J_\|$ (color) in the case of the absence of blob density profile[(a)-(d)] for the amplitude of initial current density $J_0=$ 0.25, and two $J_0$s are separated at a distance of $102\rho_s$. Here (a) $t=10/\Omega_s$, (b) $t=4000/\Omega_s$, (c) $t=8000/\Omega_s$, and (d) $t=12000/\Omega_s$.}
    \label{fig:2f_ap_jp}
\end{figure*}

\begin{figure}
    \includegraphics[width=0.99 \linewidth]{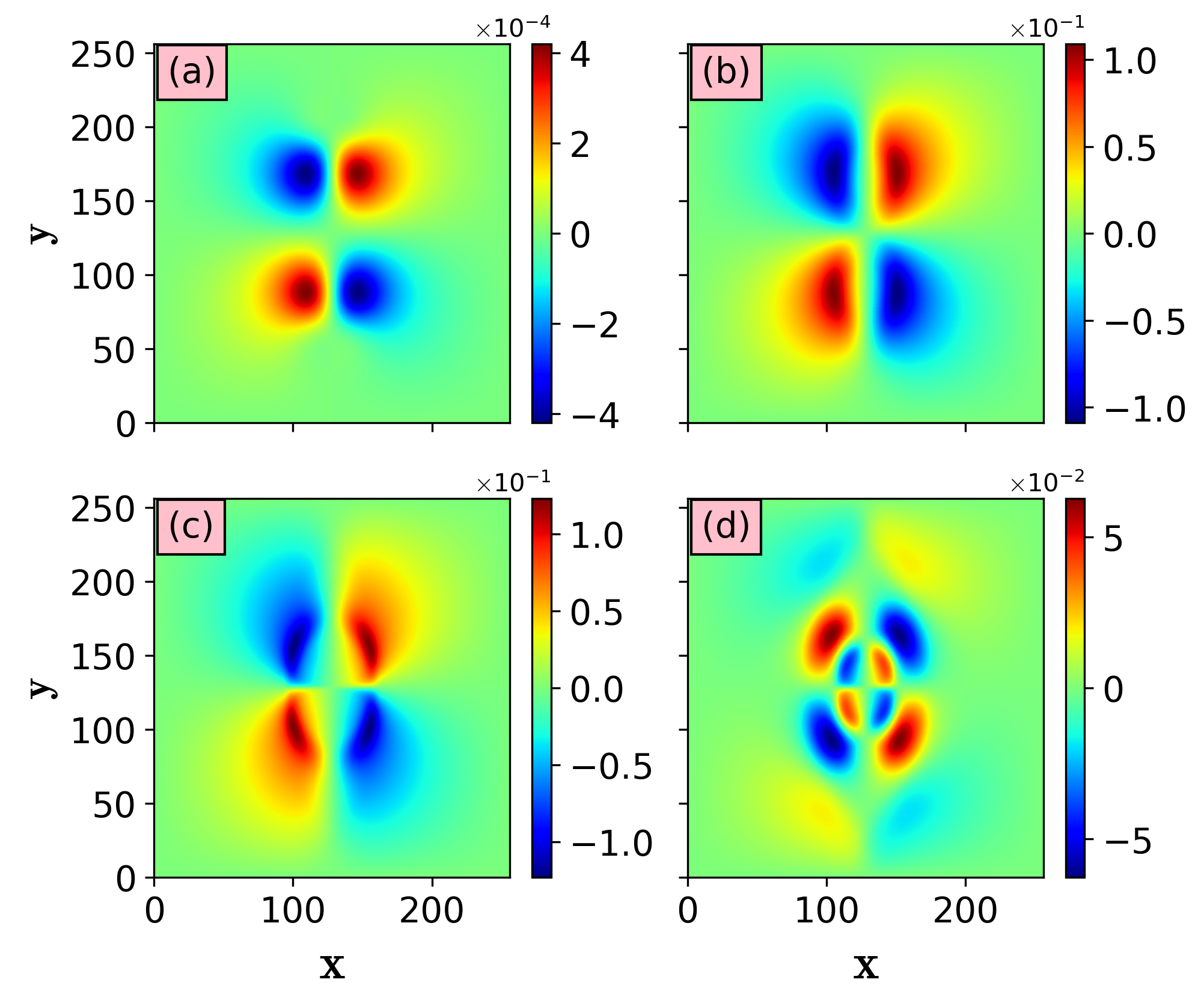}
    \caption{The time evolution of electrostatics potential in the absence of blob density profile. Here, the amplitude of initial current density $J_0=$ 0.25, and two $J_0$s are separated at a distance of $102\rho_s$; where (a) $t=10/\Omega_s$, (b) $t=4000/\Omega_s$, (c) $t=8000/\Omega_s$, and (d) $t=12000/\Omega_s$.} \label{fig:phi_2f}   
\end{figure}

Now, we will present the numerical simulation results for the case of $n_b=0$ to compare with the earlier case. 
The numerical results are obtained by solving Eqs. (\ref{eq:norm_vorticity_eq})-(\ref{eq:norm_vector_pot_eq}) considering the absence of the blob density profile ($n=1$ and $\nabla_{\perp} n=0$), where the current density was present initially. Figure \ref{fig:2f_ap_jp} illustrates the snapshot of the superposition of $A_\|$ (contour) and $J_\|$ (color). Initially, two identical Gaussian-type parallel currents having amplitude $J_0=$ 0.25 are separated by a distance of $102\rho_s$ in the poloidal direction. The attraction force between the two parallel currents brings them closer. After that, a current sheet is developed at a later time as shown in Fig.\ref{fig:2f_ap_jp}, and finally, on both sides of the current sheet, the magnetic flux surfaces reconnect, which is similar to the magnetic island coalescence phenomena in astrophysical/laboratory plasmas \cite{Coalescence_instability_finn_kaw, Pritchett_1979,D_A_Knoll_Pop_2006}. The initial (Fig.\ref{fig:2f_ap_jp}a) two small-size current densities or filaments coalesce to a bigger size (Fig.\ref{fig:2f_ap_jp}d). It is to be noted that the rotation of the current filaments is absent as the density is uniform and is independent of $r$. The motion is restricted to the $\hat{y}$ and $-\hat{y}$ direction only.\\

A snapshot of the electrostatic potential for this case is depicted in Fig.\ref{fig:phi_2f}(a)-(d). It is observed that at the initial stage ($t=10/\Omega_s$), the potential has a quadruple-like structure. The top two potential structures generate $E_x\hat{x}$ and the bottom two structures generate $-E_x\hat{x}$.  Therefore, these two opposite fields will make the current density structures come closer by $E_x\times B$ drifts. The magnitude of potential structures increases with time as seen in Fig.\ref{fig:phi_2f}(b) which creates higher $E_x$ and the two current density structures come closer at a higher velocity. Here, the $\omega$ does not depend on the density and the $\phi$ always has a quadrupolar structure as shown in Fig.\ref{fig:phi_2f}(a)-(c). That is why the two current density structures come closer in the $\hat{y}$ and $-\hat{y}$ direction without rotation as shown in Fig.\ref{fig:2f_ap_jp}. Here, after a long time ($t=12000/\Omega_s$), due to nonlinearity, the quadruple structure breaks into a complicated form as seen in Fig.\ref{fig:phi_2f}(d).\\ 

\begin{figure}
\begin{centering}
\includegraphics[width=0.90\linewidth]{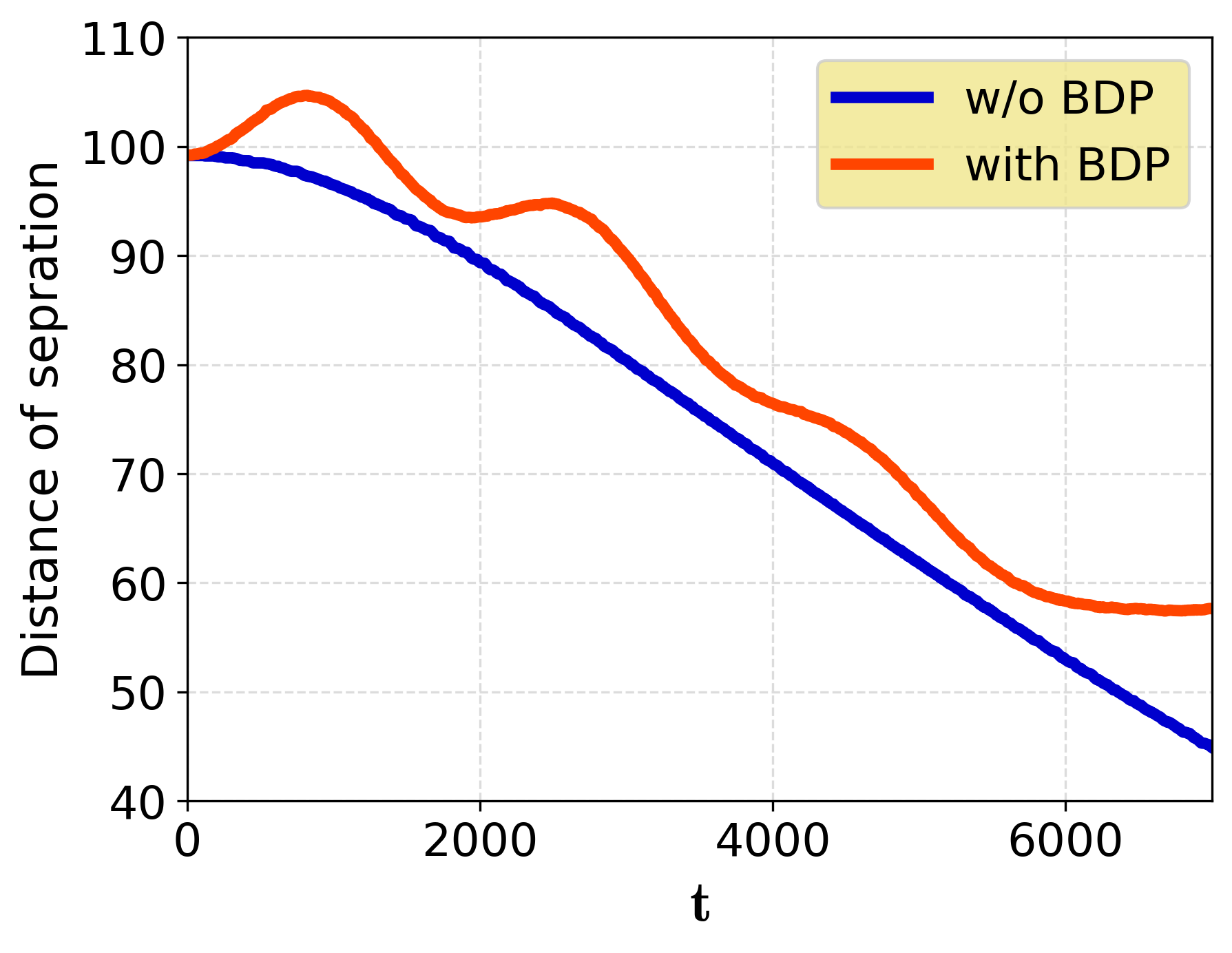} 
\par\end{centering}
\caption{The time evolution of the distance of separation between the magnetic O-points of the flux surfaces for presence (orange) and absence (blue) blob density profile. Initially, the O-points are separated at a distance of $102\rho_s$ of two different poloidal directions having an amplitude of initial current density $J_0=$ 0.25.}
\label{fig:o_point_pos_2f_3f_compare}
\end{figure}
The evolutions of the distance of separation between two blobs with time obtained in the presence and absence of blob density profile are shown in Fig.\ref{fig:o_point_pos_2f_3f_compare} for the magnitude of current density $J_0=$ 0.25. It is seen that with time, the distance of separation of blob decreases almost similarly in both cases but for the case of blob density profile, the distance of separation oscillates. This oscillation occurs because of the mutual rotation of the filaments which is not present in the case in the where the density profile is absent. This mutual rotation of the filaments is resisting to come closer. It is to be noted that the distance of separation has been measured between two O-points, where O-points have been calculated from the center of mass (COM) of each filament. The $x$ and $y$-components of the COM of each filament have been calculated using the following expression:
\begin{equation}
    x_{com}(t)=\frac{\int \int xn(t,x,y)dxdy}{\int \int n(t,x,y)dxdy},\label{eq:xcm}
\end{equation}
\begin{equation}
    y_{com}(t)=\frac{\int \int yn(t,x,y)dxdy}{\int \int n(t,x,y)dxdy},\label{eq:ycm}
\end{equation}
where the integrals in the above equation have been calculated from the whole simulation box.\\

\begin{figure}
\begin{centering}
\includegraphics[width=0.98\linewidth]{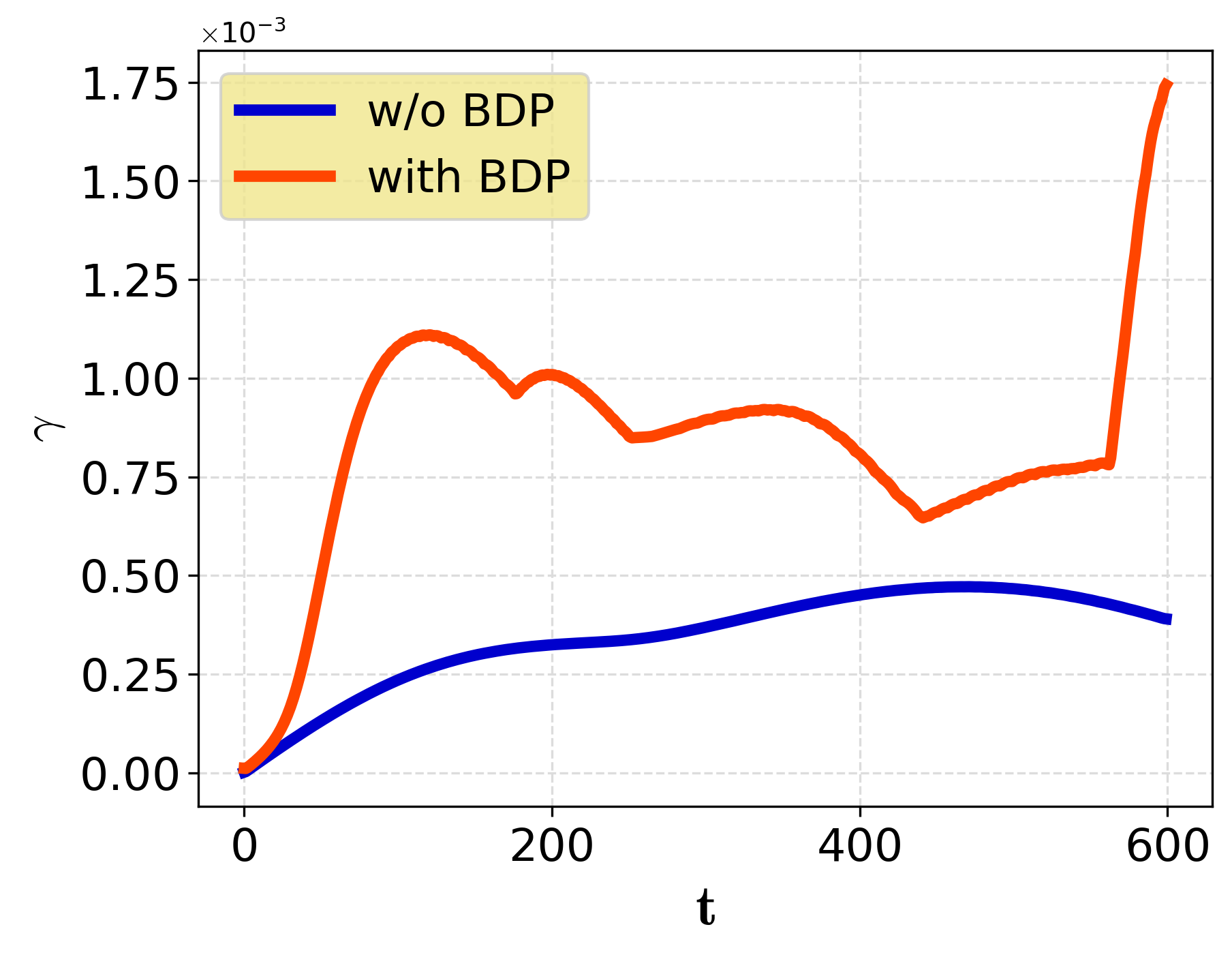} 
\par\end{centering}
\caption{The time evolution of the maximum growth of the $A_\|$ for the presence (orange) and absence (blue) of the blob density profile. Here the growth rates ($\gamma$) are calculated numerically.}
\label{fig:gr_2f_3f}
\end{figure}
The maximum numerical growth rate ($\gamma$) of the total $A_\|$ is plotted with time for with (orange) and without (blue) blob density profile as shown in Fig.\ref{fig:gr_2f_3f}. The numerical growth rate is calculated using the expression $\gamma=(1/A_\|)(dA_\|/dt)$. It has been observed that for both cases, total $A_\|$ grows with time but for the case when the blob density profile is present, the growth rate is higher than the case with the absence of blob density profile. Despite the higher growth rate, the separation distance between the filaments in the presence of the blob density profile falls more slowly with time than in the absence of the blob density profile case. This observation is mainly due to the onset of rotational motion. The rotational motion uses energy therefore it reduces the generation of magnetic field energy during the motion of the two blobs.

%-----------------------------------------------------------------------------------------------------------------------------------------------%
\subsection{The effect of blob size on merging}
%-----------------------------------------------------------------------------------------------------------------------------------------------%
\begin{figure}
\begin{centering}
\includegraphics[width=0.98\linewidth]{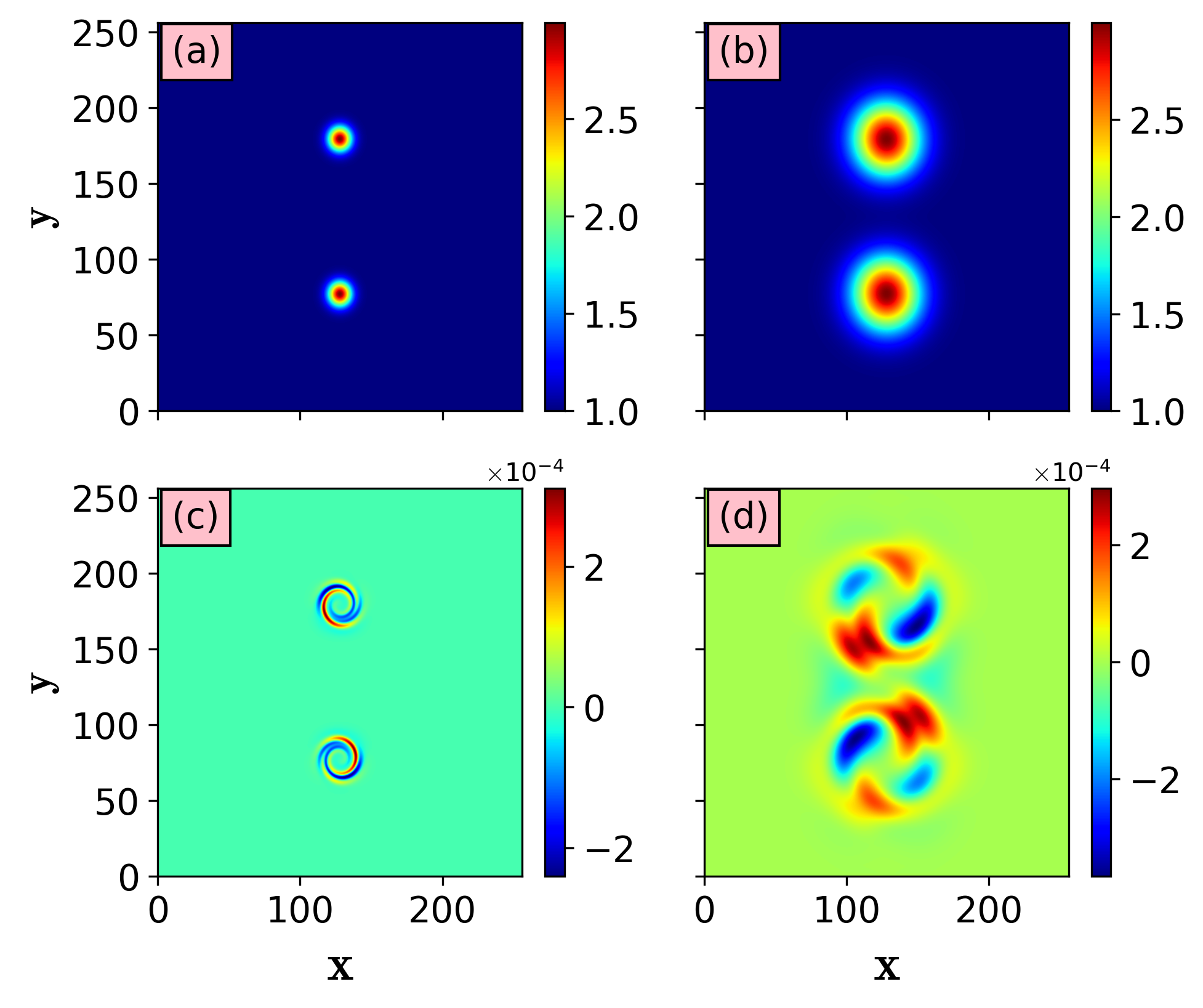} 
\par\end{centering}
\caption{Snapshot of blob for different sizes (a) $\delta=0.30$ and (b) $\delta=0.90$, and the corresponding vorticity at time $t=1000/\Omega_s$. Here, the amplitude of initial current density $J_0=$ 0.25, and two $J_0$s are separated at a distance of $102\rho_s$ with $n_b=2$.}
\label{fig:n_vort_delta}
\end{figure}

In the previous subsection, we examined the behavior of filaments in the presence and absence of a blob density profile. It was found that filaments rotate/spin around their axis when the blob density profile is present, but this rotation/spin is absent when there is no blob density profile.\\

\begin{figure}
\begin{centering}
\includegraphics[width=0.98\linewidth]{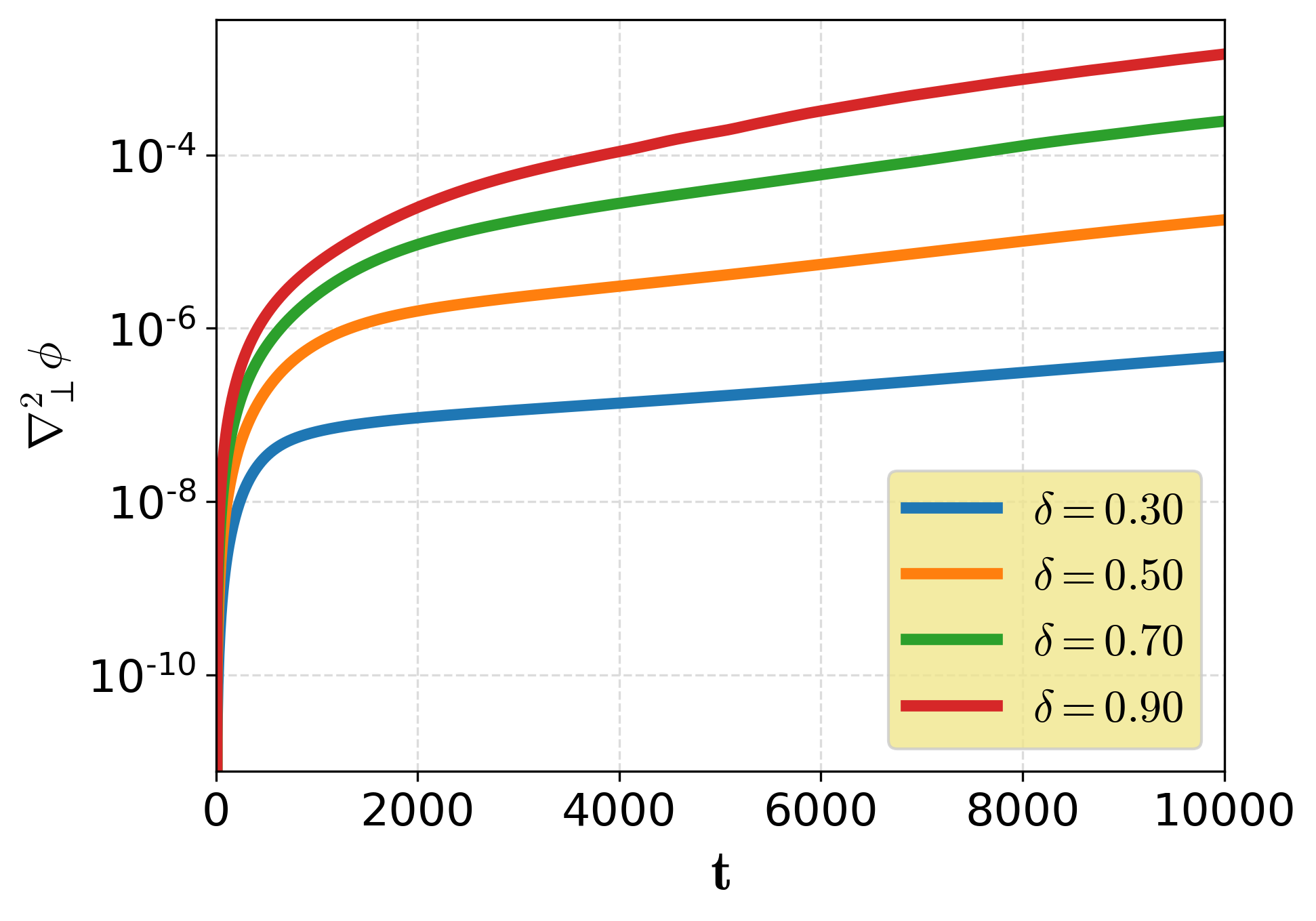} 
\par\end{centering}
\caption{Time evolution of volume averaged vorticity for different sizes. Initially, the blobs are separated at a distance of $102\rho_s$ of two different poloidal directions having an amplitude of initial current density $J_0=$ 0.25 with $n_b=2$.}
\label{fig:average_vort}
\end{figure}

In this section, we will examine the impact of varying sizes of a blob on the dynamics of a plasma blob in the presence of BDP. Initially, we seeded two identical blobs with a separation distance of $102\rho_s$ in the $y$-direction. Figures \ref{fig:n_vort_delta}(a)-(b) show the snapshot of the plasma blob for (a) $\delta=0.30$ and (b) $\delta=0.90$ at time $t=1000/\Omega_s$. Figures \ref{fig:n_vort_delta}(c)-(d) illustrate the vorticity at $t=1000/\Omega_s$ for the aforementioned sizes of the blob, respectively. The rotational movement of the blob structure is not fully discernible in Figs.(a) and (b) of Fig.\ref{fig:n_vort_delta}. However, in Figs.(c) and (d), the vorticity structure reveals an asymmetrical interaction of its pole in both circumstances. It is confirmed that the rotation or spin is always present in the blob structure when the BDP is present. Figures \ref{fig:n_vort_delta}(c)-(d) demonstrate that the difference between the maximum negative and maximum positive values of vorticity is greater for the larger blob ($\delta=0.90$) compared to the smaller blob ($\delta=0.30$). The larger blob exhibits a higher rotational speed. To verify this, we conducted simulations using four different blob sizes while keeping all other parameters constant. Figure \ref{fig:average_vort} shows the vorticity averaged over volume as a function of time. The initial vorticity of all the blobs is zero, but it grows over time. Based on this observation, we can conclude that the rotation speed increases as the blobs approach each other. Again, the magnitude of the vorticity is higher for the bigger blob than the smaller blob, resulting in a higher rotation for the larger blob. The calculated rotation speed at $t=1000/\Omega_s$ for the $\delta=0.30$ and $\delta=0.90$ are $0.00042s_s$ and $0.002c_s$ respectively.  \\

\begin{figure}
\begin{centering}
\includegraphics[width=0.98\linewidth]{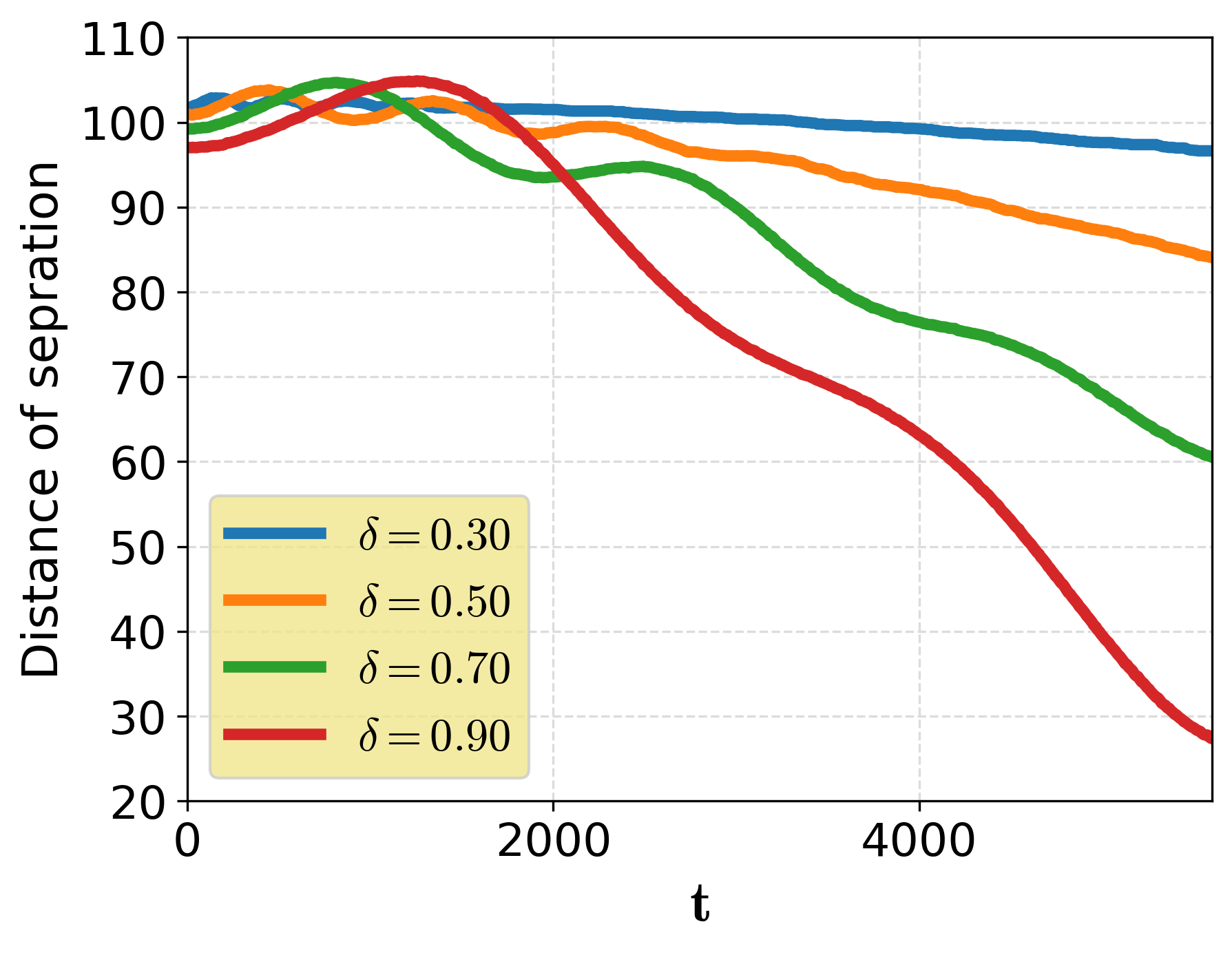} 
\par\end{centering}
\caption{The time evolution of the distance between the COM position of the blob for different sizes ($\delta$). Initially, the blobs are separated at a distance of $102\rho_s$ with an amplitude of initial current density $J_0=$ 0.25 with $n_b=2$.}
\label{fig:o_pos_delta}
\end{figure}

\begin{figure}
\begin{centering}
\includegraphics[width=0.98\linewidth]{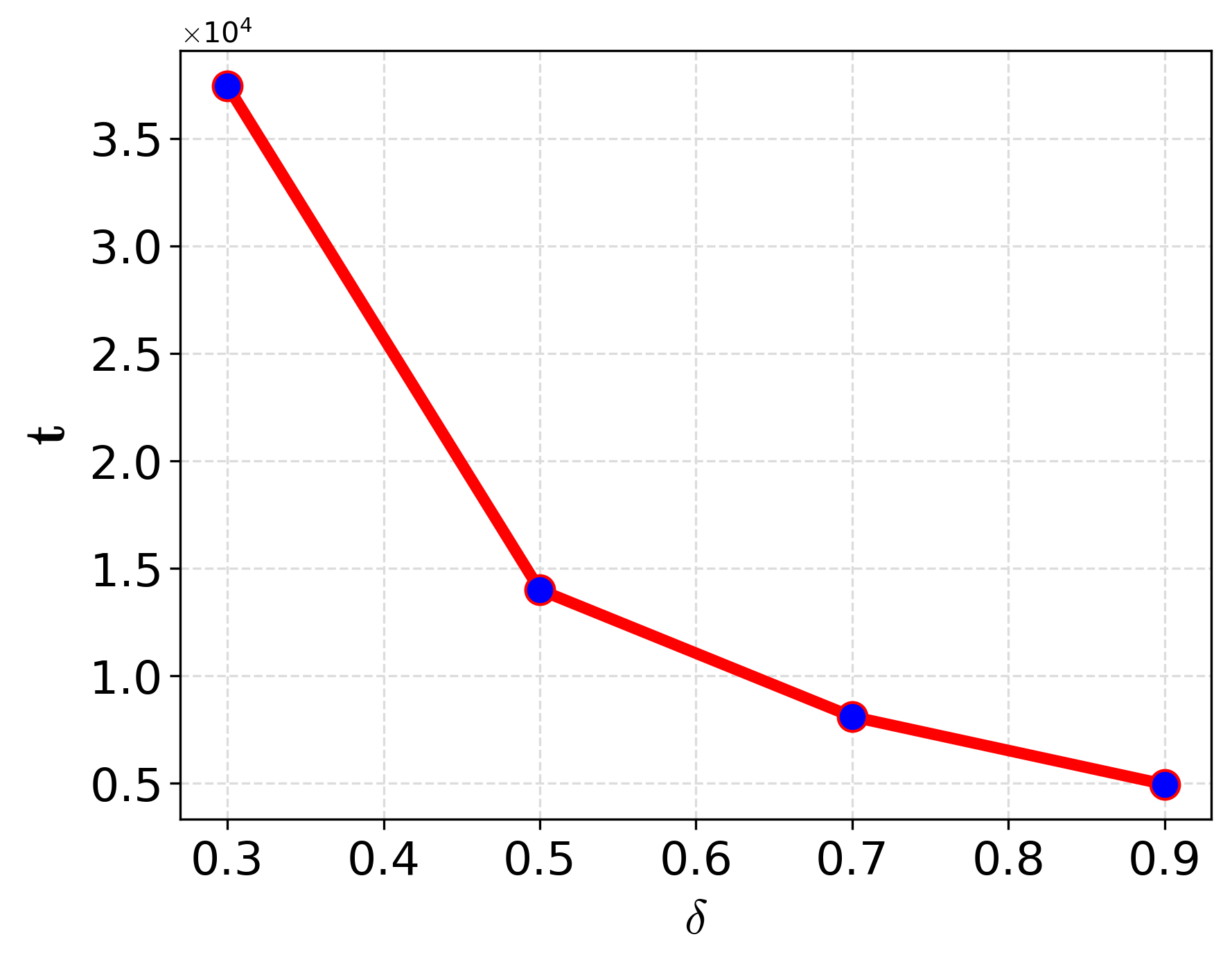} 
\par\end{centering}
\caption{Plot for the time taken for $1/e$-th of the initial distance of separation of the blobs with different sizes. Initially, the blobs are separated at a distance of $102\rho_s$ of two different poloidal directions having an amplitude of initial current density $J_0=$ 0.25 with $n_b=2$.}
\label{fig:time_delta}
\end{figure}

Figure \ref{fig:o_pos_delta} shows the time evolution of the distance between the COM position of the blobs (distance of separation) for different sizes ($\delta$). Initially, the blobs are located at two different poloidal directions that are separated by a distance of $102\rho_s$. The amplitude of their initial current density is $J_0=$ 0.25 with $n_b=2$. It is observed that for the larger-sized blobs, the distance of separation decreases at a higher rate than that for the smaller blobs. This is also shown in  Fig.\ref{fig:time_delta}, where we plot the time taken to  $1/e$-th of the initial distance of separation between the blobs for different sizes and observe that the bigger blob takes less time than the smaller blob. From Eq.\ref{eq:A5}, the acceleration between the blob is proportional to the square of the blob size ($\delta$). So the acceleration will be higher for the bigger blob than the smaller one which is consistent with the analytical calculation.

%-----------------------------------------------------------------------------------------------------------------------------------------------%
\subsection{The effect of $J_0$ on merging}
%-----------------------------------------------------------------------------------------------------------------------------------------------%
\begin{figure}
    \includegraphics[width=0.99\linewidth]{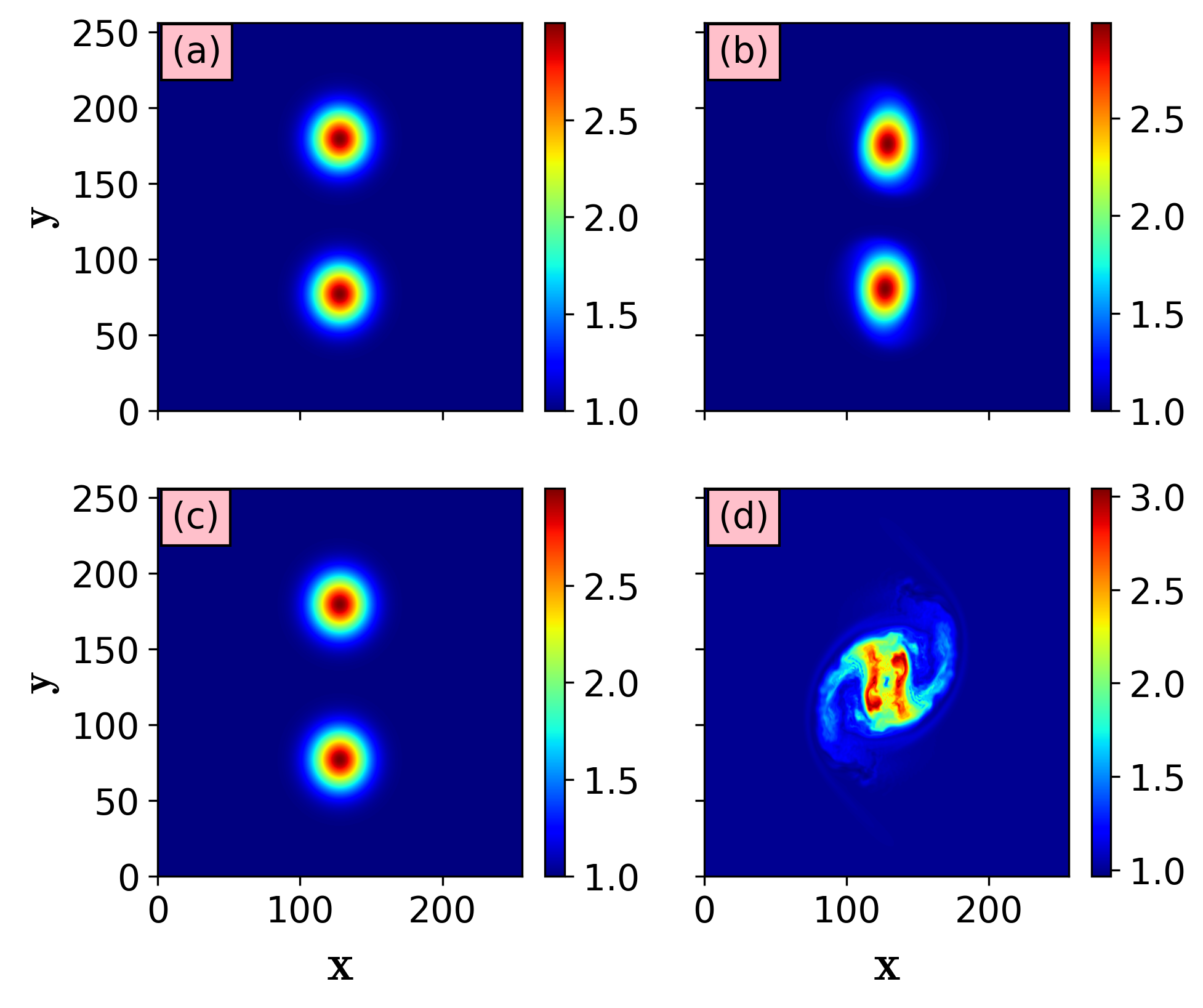}
    \caption{The time evolution density for the initial parallel current density $J_0=$ 0.125 (a and b) and $J_0=$ 0.75 (c and d), respectively. Initially, we set $n_b=2$. Here $t=10/\Omega_s$ (a and b) and $t=4500/\Omega_s$ (c and d). }
    \label{fig:n_ap_jp_3f_compare}
\end{figure}
\begin{figure}
\begin{centering}
\includegraphics[width=0.98\linewidth]{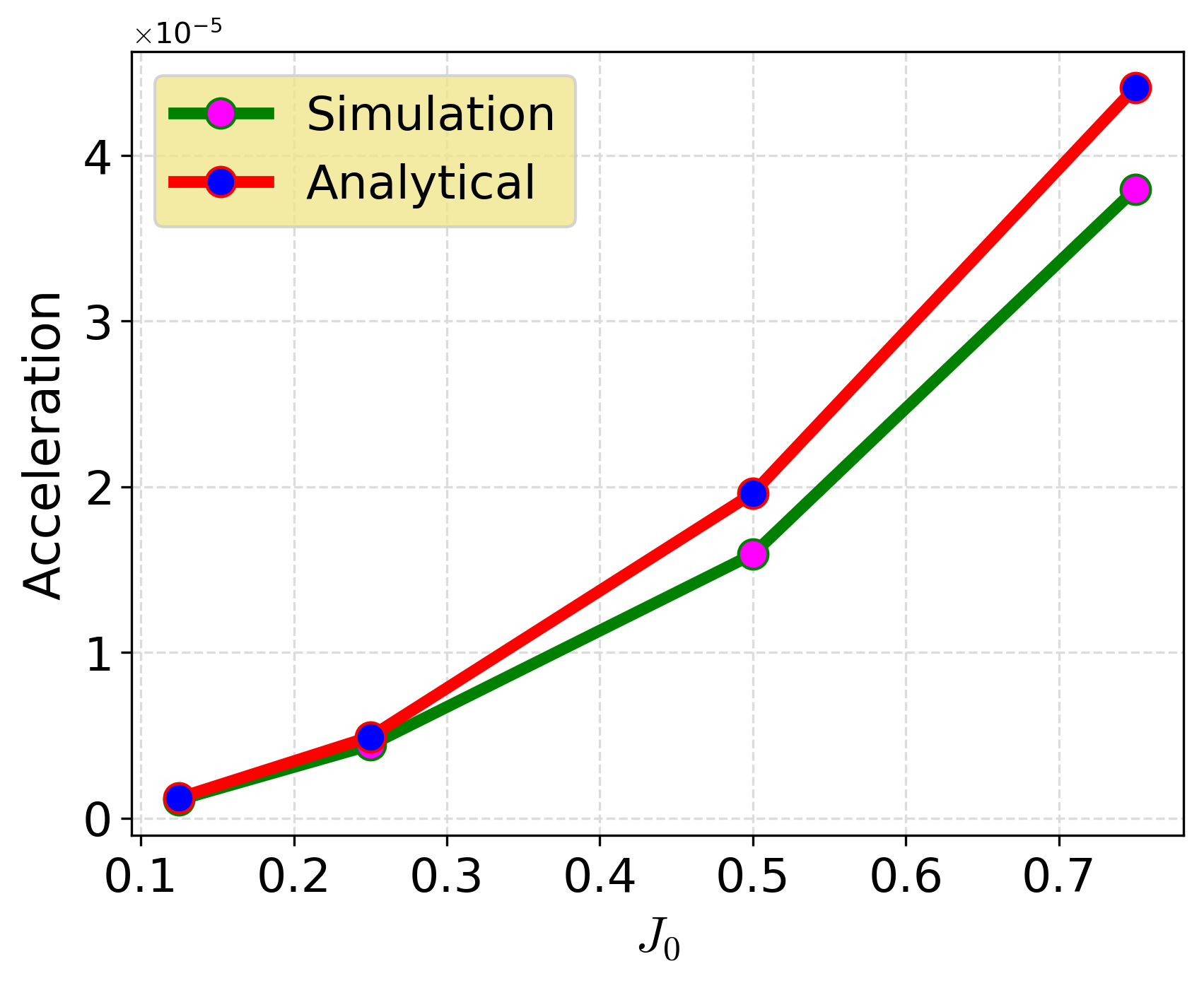} 
\par\end{centering}
\caption{Acceleration in the normalized unit (1 normalized unit=$2.64\times10^{15}$ cm/s$^2$) of the filament obtained from analytical and simulation data for different $J_0$ with $n_b=2$ at time $t=0$.}
\label{fig:acceleration}
\end{figure}

 
 We next discuss the effect of varying the amplitude of the parallel current density on the merging process when $n_b$ is kept the same. For comparison, we present the results for $J_0=$ 0.125 and 0.75 in Figs.\ref{fig:n_ap_jp_3f_compare}(a)-(d). Figures (a)-(b) show the snapshot of plasma density for $J_0=$ 0.125 at time $t=10/\Omega_s$ and $t=4500/\Omega_s$. Figures \ref{fig:n_ap_jp_3f_compare}(c)-(d) represent the same with $J_0=$ 0.75 at the above two instants of times. Here it is observed that in the case of $J_0=$ 0.75, the blobs come closer than in the case of $J_0=0.125$. Additionally, there is rotation in the blobs (Fig.\ref{fig:n_ap_jp_3f_compare}(d)) as is expected for finite $n_b$. Again, two blobs are initially accelerated because of the Lorentz force acting on them. The acceleration of the blobs due to the Lorentz force can be explained from Eq.(\ref{eq:A5}) and is proportional to the $J_\|^2$. To show the magnitude of the acceleration as a function of $J_0$ we have plotted the magnitude of the acceleration for $J_0=$ 0.125, 0.25, 0.5, and 0.75 and marked them as magenta circle dots in Fig.\ref{fig:acceleration}. The magnitude of the acceleration is compared with the analytical estimation given in Eq.(\ref{eq:A5}) and shown as blue-filled circle marks. It is seen that the analytical and simulation results are close to each other. It is to be noted that for higher $J_0$, the simulation results are deviated from the analytical one. This could be because of the nonlinear effect of the system that is not given much attention in this present work. \\

\begin{figure}
\begin{centering}
\includegraphics[width=0.98\linewidth]{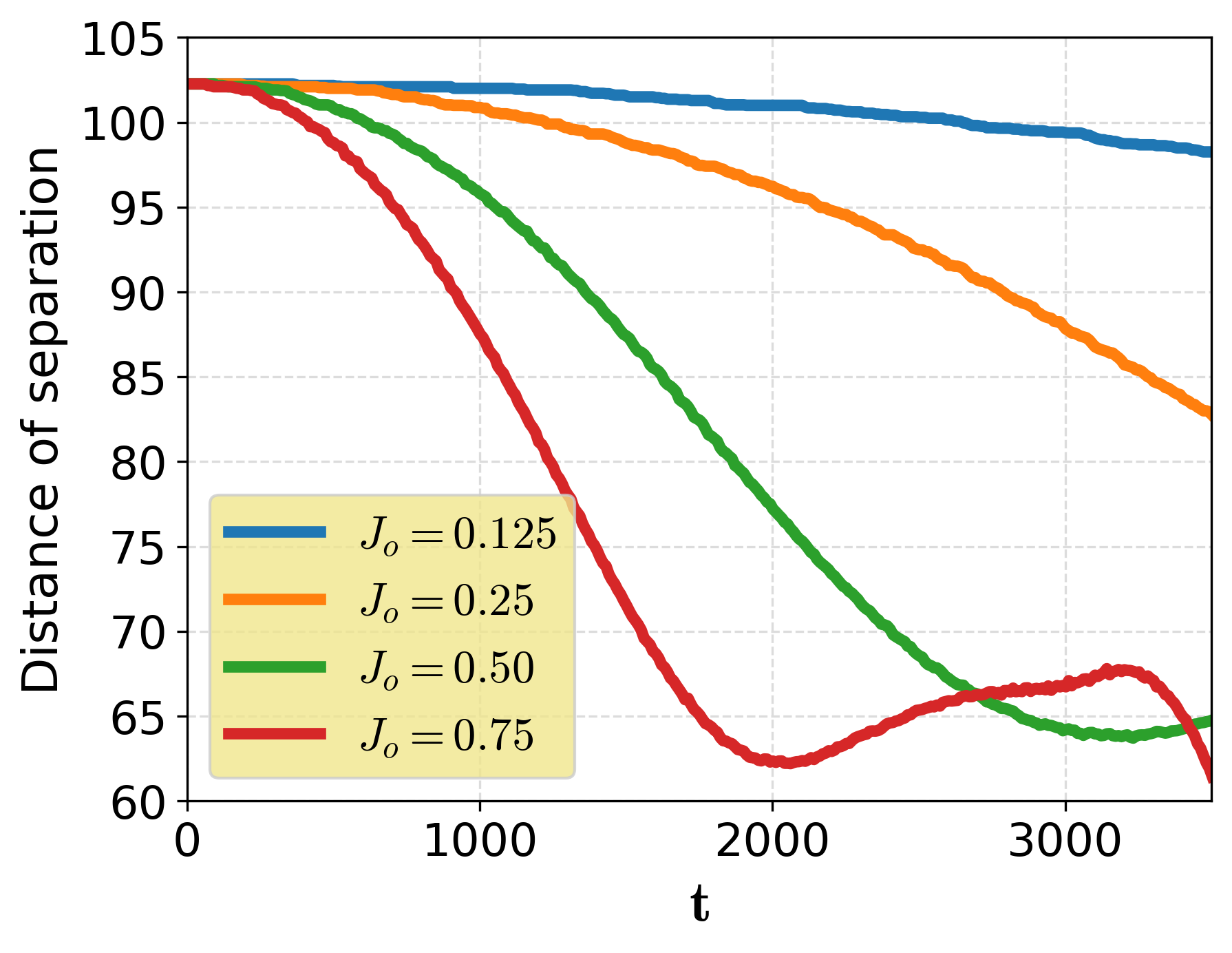} 
\par\end{centering}
\caption{The plot shows the time evolution of the distance of separation of the COM positions for the blobs with finite $n_b$  with an initial distance of separation of $102\rho_s$ in the $y$-direction for four different values of $J_0=$ 0.125, 0.25, 0.5, and 0.75.}
\label{fig:o_pos_0.4_all_3f_compare}
\end{figure}

Furthermore, Fig.\ref{fig:o_pos_0.4_all_3f_compare} shows the separation distance of the COM positions as a function of time for the different initial amplitudes of $J_0$. It is observed that, for lower $J_0$ (0.125 and 0.25), the distance decreases very slowly. On the other hand, for higher $J_0=$ 0.5, and 0.75, the distance of separation decreases rapidly to indicate a higher acceleration for higher $J_0$. An oscillation in the separation distance is observed for $J_0=$ 0.5 and 0.75. A similar observation related to the oscillation has been reported in Ref. \cite{D_A_Knoll_Pop_2006}.  

%-----------------------------------------------------------------------------------------------------------------------------------------------%
\subsection{The effect of $n_b$ on merging}
%-----------------------------------------------------------------------------------------------------------------------------------------------%

\begin{figure}
\includegraphics[width=0.99\linewidth]{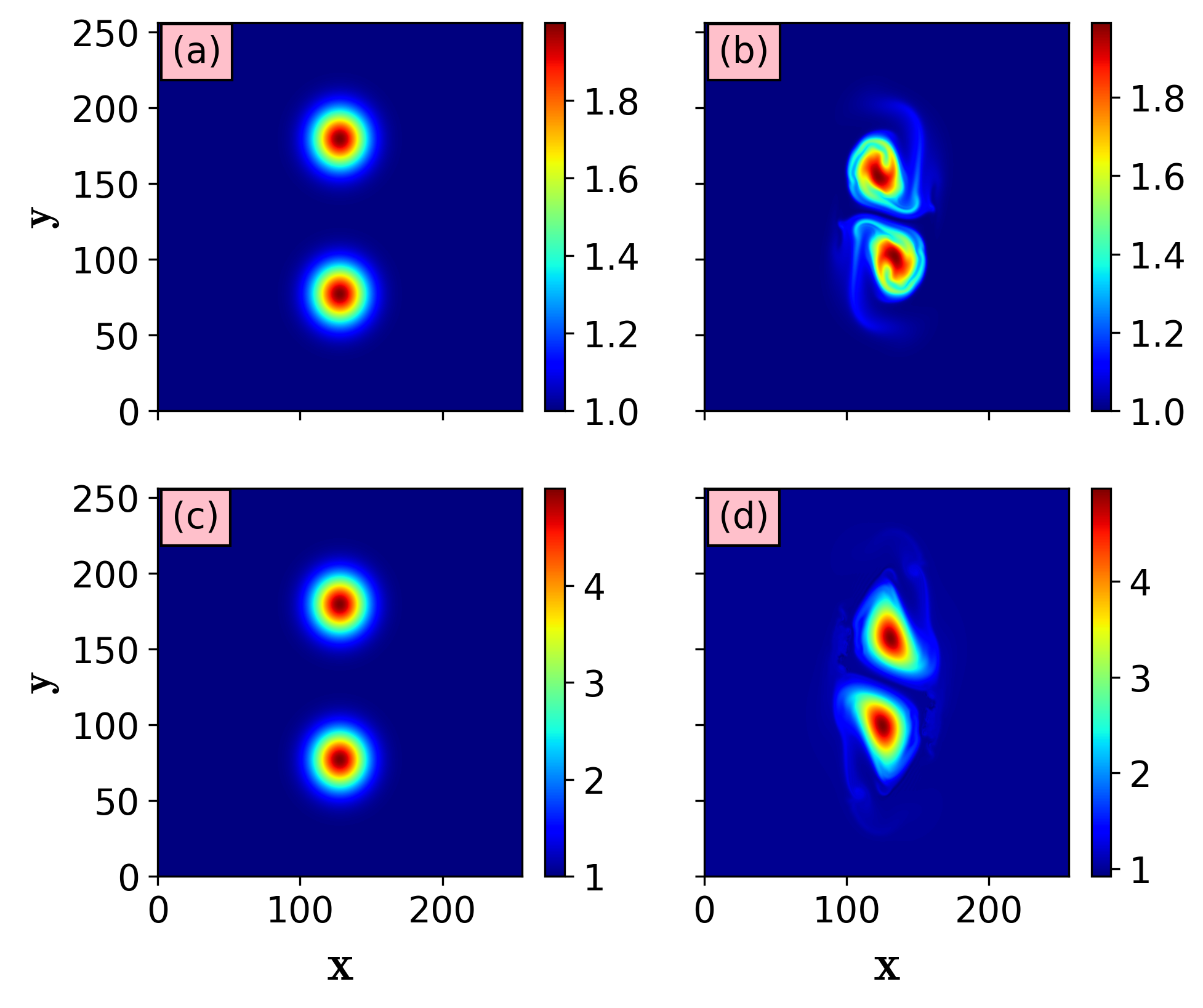}.
\caption{The time evolution density for the different magnitudes of blob density $n_b$=1 (a and b) and $n_b$=4 (c and d), respectively with initial $J_0=$ 0.25. Here $t=10/\Omega_s$ (a and b) and $t=6500/\Omega_s$ (c and d).}
\label{fig:n_ap_jp_n_compare}
\end{figure} 
In the previous subsection, we discussed the dynamics of the filaments in the presence of the blob density profiles for different $J_0$ with  $n_b=2$. Here we will investigate the impact of $n_b$ on the blob dynamics.  For this purpose, we have varied $n_b$  from $n_b=1$ to $n_b=4$ with an initial distance of separation $102\rho_s$ and $J_0=0.25$. Figures \ref{fig:n_ap_jp_n_compare}(a)-(b), show snapshots of the density blob evolution for the blob magnitude $n_b=1$ at time $t=10/\Omega_s$ and $t=6500/\Omega_s$, respectively. It is observed that the density blobs are coming closer to each other with a rotation. Figures.\ref{fig:n_ap_jp_n_compare}(c)-(d) show the density of blobs with $n_b=4$. Figure \ref{fig:n_ap_jp_n_compare}(d) also shows the rotation but the poloidal movement of blobs is slower. We have compared the rotation speeds obtained from $\vec{E}_r\times \vec{B}$ drift. These are $0.0024c_s$ and $0.001c_s$ at $t=6500/\Omega_s$ for  $n_b=4$ and $n_b=1$, respectively.  The higher rotation speed for the higher blob plasma density has been attributed to Eq.\ref{eq:ln_condition} It is to be noted that the radial gradient of $n$ is higher for a higher $n_b$ for the fixed size of the filament, therefore, it rotates faster.\\

\begin{figure}
\begin{centering}
\includegraphics[width=0.98\linewidth]{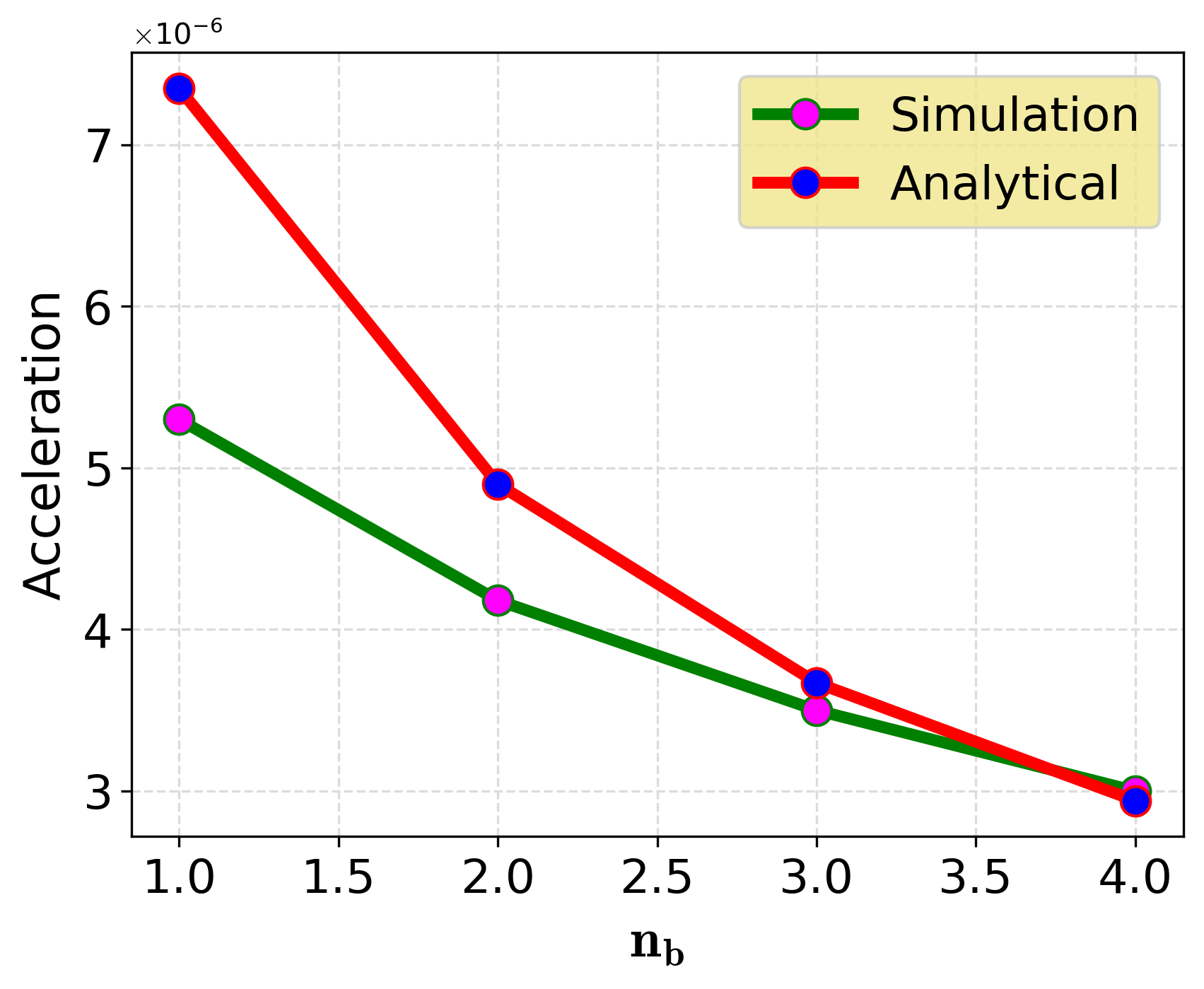} 
\par\end{centering}
\caption{Acceleration in the normalized unit (1 normalized unit=$2.64\times10^{15}$ cm/s$^2$) of the blob obtained from analytical and simulation data for different $n_b$ values. Here, initially, we set $J_0=$ 0.25.}
\label{fig:acceleration_n}
\end{figure}

\begin{figure}
\begin{centering}
\includegraphics[width=0.98\linewidth]{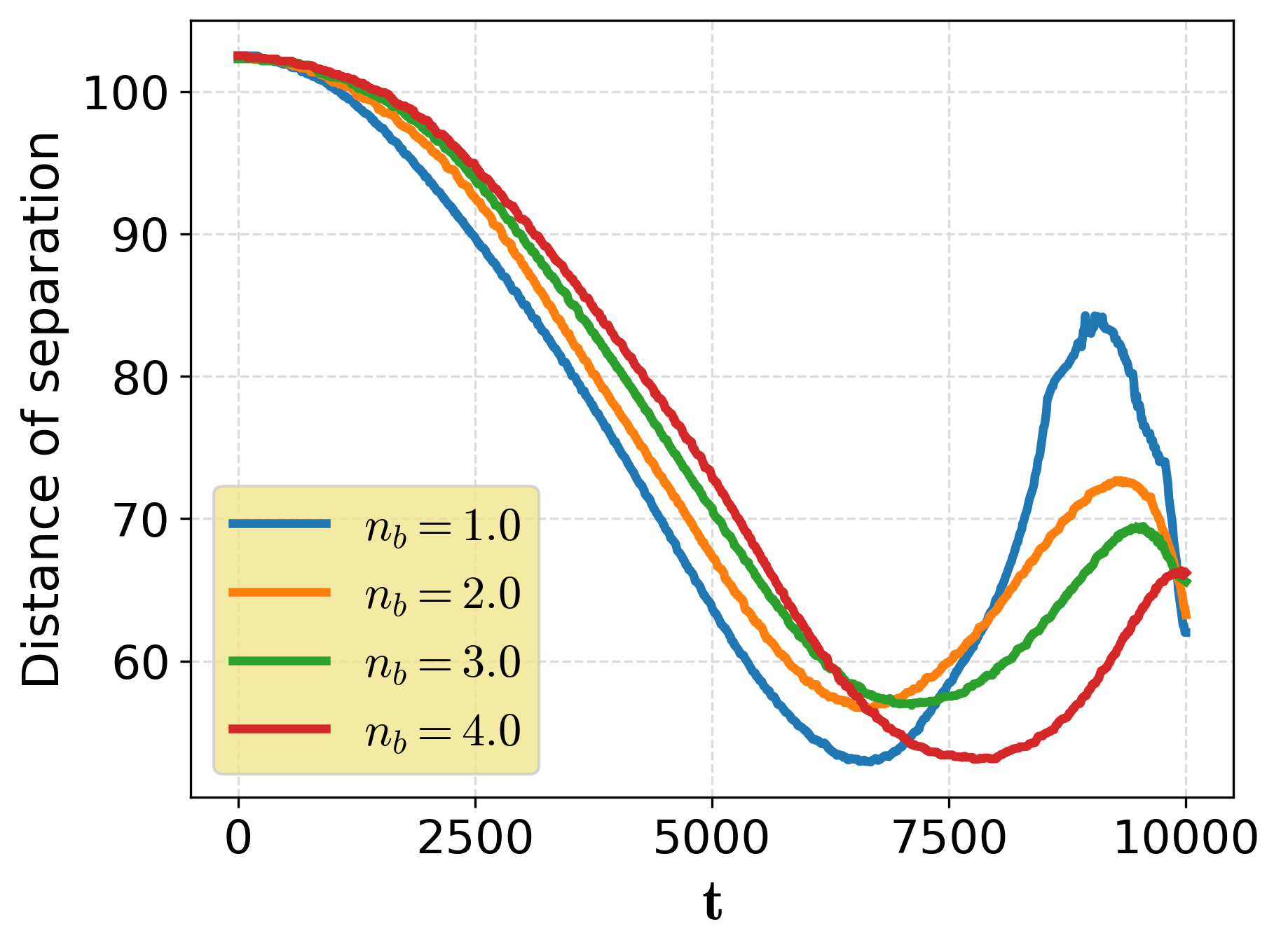} 
\par\end{centering}
\caption{The time evolution of the separation distance between the COM positions for  $n_b$=1, 2, 3, and 4.}
\label{fig:o_pos_0.4_all_3f_compare_n}
\end{figure}

The acceleration of the blob is also calculated from the simulation and analytical data as shown in Fig.\ref{fig:acceleration_n} which shows that it decreases with $n_b$ and both cases match closely. This type of behavior is shown in Eq.(\ref{eq:A5}) analytically.  Again, the separation distances of the blob for the different $n_b$ have been plotted in Fig.\ref{fig:o_pos_0.4_all_3f_compare_n} using $J_0=0.25$. The distance of separation is calculated using the COM positions between the filaments as a function of time. It is observed that the separation distance between the blob also decreases with time and it is similar for all $n_b$ values but for lower $n_b$, it falls at higher rates. This is because the acceleration is higher for the lower $n_b$ as per Eq.(\ref{eq:A5}). The sloshing phenomena (oscillation in the distance of separation with time) are also seen.

%-----------------------------------------------------------------------------------------------------------------------------------------------%

\section{Discussion and Conclusions} \label{sec:discussion}

%-----------------------------------------------------------------------------------------------------------------------------------------------%
In this study, we have investigated the dynamics of plasma blobs in the SOL regions of a tokamak. We have used a high $\beta$ plasma where plasma parameters in the SOL region are similar to ITER.  The dynamics of the blob have been studied in the presence and absence of a density profile in the blob composition.  In the absence of a blob density profile, the dynamics of the current filaments are mainly governed by the coalescence instability. In the presence of a blob density profile, the plasma blob rotates about its axis along the magnetic field direction. The rotation is related to the density scale length of the blob profile.\\

The two plasma blobs attract each other by the Lorentz force. The acceleration due to the attraction is proportional to $J_0^2$ and inversely proportional to $n_b$ that have been demonstrated numerically.  We have also calculated the distance of separation between two blobs as a function of time for a given $J_0$, and $n_b$.  The rotation of the plasma blob has been studied with $n_b$ where the rotational speed increases with the density gradient at the plasma blob. The distance of separation for the higher $J_0$ oscillates similar to Sloshing \cite{D_A_Knoll_Pop_2006}.\\

However, in our present model, we have neglected the force due to effective gravity, which can cause a radial motion of plasma blobs. We hope to investigate this in the future to study its impact on the blob dynamics in the SOL region and contrast it with the effects studied in the present paper. 

\begin{acknowledgments}
The simulations were performed on the Antya cluster at the Institute for Plasma Research (IPR). This work has been carried out as part of the PhD work of the first author who is registered with HBNI, Mumbai. He conceptualized the problem and constructed the code using BOUT++, and he is the primary contributor to the work. A. Sen is grateful to the Indian National Science Academy (INSA) for the INSA Honorary Scientist position.
\end{acknowledgments}

\bibliographystyle{unsrt}
\bibliography{citation}

\end{document}